\documentstyle[twocolumn,aps,epsfig, floats,times]{revtex}

\draft \tighten
\newcommand{\ket}[1]{\left | \, #1 \right \rangle}

\title{Experimental observation of  four-photon entanglement from
 down-conversion}
\author{Manfred Eibl$^{1}$, Sascha Gaertner$^{1,2}$, Mohamed Bourennane$^{1,2}$,
Christian Kurtsiefer$^{2}$, Marek \.Zukowski$^3$  and Harald Weinfurter$^{1,2}$}

 \address{$^1$ Max-Planck-Institut f\"ur Quantenoptik, D-85748 Garching, Germany}
 \address{$^2$ Sektion Physik, Ludwig-Maximilians-Universit\"at, D-80799 M\"unchen, Germany}
 \address{$^3$Instytut Fizyki Teoretycznej i Astrofizyki Uniwersytet Gda\'nski, PL-80-952 Gda\'nsk, Poland}
\date{\today}

\begin{document}

\maketitle

\begin{abstract}
We observe  polarization-entanglement between four photons
produced from a single down-conversion source. The non-classical correlations
between the measurement results violate a generalized Bell inequality for four
qubits. The characteristic properties and its easy generation with high
interferometric contrast make the observed four-photon  state well-suited for
implementing advanced quantum communication schemes such as multi-party
quantum key distribution, secret sharing and telecloning.
\end{abstract}


\vspace{1cm}

Entanglement between more than two particles is the key ingredient
for advanced multi-party quantum communication. A number of
proposals, e.g.  telecloning\cite{telecloning},  reduction of
the communication complexity \cite{commcompl}, or secret
sharing\cite{secrshare} utilize multi-particle entanglement for
quantum communication protocols.

Only few experiments have demonstrated entanglement between more
than two qubits. Whereas the strong coupling between atoms enables
engineered state preparation \cite{entatoms} as required for
quantum computation, entangled multi-photon states are best
suited for communication purposes. Interference of independently
created photon pairs was used together with conditional detection 
for the first observations of three- and four-photon GHZ states
\cite{3and4GHZexp}.
Yet, those schemes required  fragile  interferometric setups
which limit their applicability and make detailed investigations
difficult. Moreover, it is important to generate also different types of
entangled  multi-photon states required for quantum communication.

In this letter, we show that a polarization-entangled four-photon state can be
directly observed behind a single pulsed spontaneous parametric
down-conversion (SPDC) source.
In contrast to previous techniques, the state forms without overlapping
photons at beam splitters.  The observed entanglement can be used for 
telecloning and, as shown here, for multi-party key distribution
and secret sharing.

In spontaneous parametric down-conversion, there is a reasonable probability
of simultaneously producing four photons for strong pump power. However, if
the type-II 
down-conversion \cite{poentDC} is adjusted to give  polarization 
entanglement for a pair emitted into the two spatial modes
$a_0$ and $b_0$, the state of the four photons emitted into these
modes is not simply the product of two entangled pairs
\cite{proposal}.
Due to their bosonic nature, the emission of two otherwise indistinguishable
photons with identical polarization into the same direction is twice
as probable as the emission of two photons with orthogonal polarization.
\cite{otherdegfreed}.
Splitting each of the two
modes at a non-polarizing beam splitter enables the observation of
correlations due to the entanglement between four photons.

\begin{figure}
\begin{center}
\epsfig{file=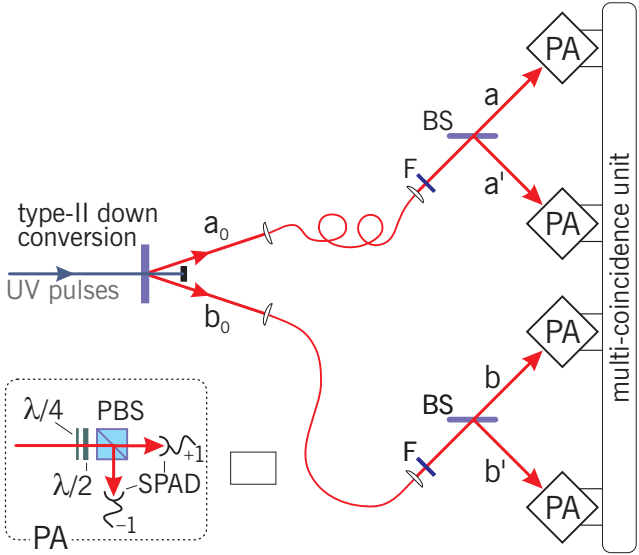,width=8.5cm}
\end{center}
\caption{\label{setupfig}Experimental setup. The four
photons are emitted from the BBO crystal (type-II phase matching) into two spatial modes $a_0$ and $b_0$, and
distributed into the four modes $a, a', b, b'$ by 50-50
beam splitters (BS) behind interference filters (F). To
characterize the resulting four-photon state $\ket{\Psi^{(4)}}$,
polarization analysis (PA) in various bases is performed for each
mode using $\lambda/4$ and $\lambda/2$ wave plates in front of
polarizing beam splitters (PBS) and single photon avalanche
detectors (SPAD). Joint photodetection events in the four arms are
recorded in a multi-coincidence unit. }
\end{figure}

In our experiment we select events such that one photon is detected in
each of the four outputs ($a$, $a'$, $b$, and $b'$) of
the beam splitters (Fig.~\ref{setupfig}). These four-photon coincidences can be
explained with the four-photon state

\begin{eqnarray}
&&\ket{\Psi^{(4)}}  =  \sqrt{\frac{1}{3}}
 \left(  \ket{HHVV}+ \left. \ket{VVHH} \right. \right. \label{state}\\
  & & - \frac{1}{2}\left.
 \left( \ket{HVHV}-\ket{HVVH}-\ket{VHHV}+\ket{VHVH}\right) \right)\,,
\nonumber
\end{eqnarray}
where the four entries in the state vectors indicate horizontal ($H$) or
vertical ($V$) polarizations of the photons in arm $a, a', b$ and $b'$.
This state is a superposition of a four-photon
Greenberger-Horne-Zeilinger (GHZ) state and a product of two
Einstein-Podolsky-Rosen (EPR) pairs,
\begin{eqnarray}
\ket{\Psi^{(4)}} & = &
\sqrt{\frac{2}{3}}\ket{GHZ}_{aa'bb'}-\sqrt{\frac{1}{3}}\ket{EPR}_{aa'}\ket{EPR}_{bb'}
, \label{stateGHZ}
\end{eqnarray}
 where the GHZ state is equal to
$$\frac{1}{\sqrt{2}}(\ket{HHVV}_{aa'bb'}+\ket{VVHH}_{aa'bb'})$$ and
the EPR state is the so-called Bell state
$$\ket{\Psi^-}=\frac{1}{\sqrt{2}}(\ket{HV}_{xx'}-\ket{VH}_{xx'})$$
with $x=a,b$.
 Due to the linearity of the beam splitters,  for
this state the detection of two $H$-polarized photons, for example in the arms
$a$ and $a'$, has the same probability as all
possible combinations of having orthogonally polarized photons in
these arms.

The generic form of state $\ket{\Psi^{(4)}}$ (eq.~\ref{stateGHZ}) is
invariant under {\em identical} 
basis changes by the four observers, i.e., 
$\ket{\Psi^{(4)}}$ remains a superposition of a four-photon GHZ state and a
product of two EPR pairs when all
four observers use identically, but otherwise arbitrarily oriented
polarization analyzers. This contrasts with the GHZ 
states which loose their characteristic two-component form under
such a basis change. This feature is  related to the
fact that the four-photon state of equation (\ref{state}) can be
viewed as the result of cloning an EPR pair. It is known that quantum
cloning cannot be perfect. Thus, a four-photon GHZ contribution
is created in addition to the product of two
EPR pairs \cite{cloning}. Nevertheless, the property of
maintaining the characteristic form under basis transformation is
carried over from the EPR singlet state to its quantum clone
$\ket{\Psi^{(4)}}$.

{ To  observe this four-photon entangled state, it is necessary to select
single spatial modes and to erase the possible frequency
correlations of the original photon pairs. This can be achieved by using
{\em pulsed} parametric down conversion and by detecting the photons behind
narrow-bandwidth filters, resulting in a coherence time longer than the pump
pulse duration. \cite{ZZW}. }

In our experiment we used the UV-pulses of a frequency-doubled
mode-locked Ti:Sapphire laser (pulse length 180~fs) to pump SPDC in a 2~mm
thick properly oriented  BBO crystal at a center wavelength of 390~nm. The
pump beam was 
focused to a waist of 100~$\mu$m inside the crystal, and the
repetition rate was 76~MHz with an average power of 450~mW. The
degenerate down-conversion emission into the two characteristic
type-II crossing directions was coupled into single mode optical
fibers to exactly define the spatial emission modes.
Behind the fibers the down-conversion light passed 
interference filters ($\Delta\lambda=3$~nm), and was split at
dielectric 50\%-50\% beam splitters into four distinct spatial
modes. Polarization analysis in each of the four outputs behind
the beam splitters was performed by a combination of quarter- and
half-wave plates together with polarizing beam splitters. The four
photons were detected by single photon Si-avalanche diodes, and
registered with an eight-channel multi-coincidence unit. This
analysis system recorded every possible coincidence between the
eight detectors, and thus allowed efficient registration of the 16
relevant fourfold coincidences.
The eight detectors exhibit different efficiencies due to production
tolerances. If not stated otherwise, the
rates presented here are therefore corrected for the separately calibrated
efficiencies, and the errors given are deduced from propagated
Poissonian counting statistics of the raw detection events.

\begin{figure}
\begin{center}
\epsfig{file=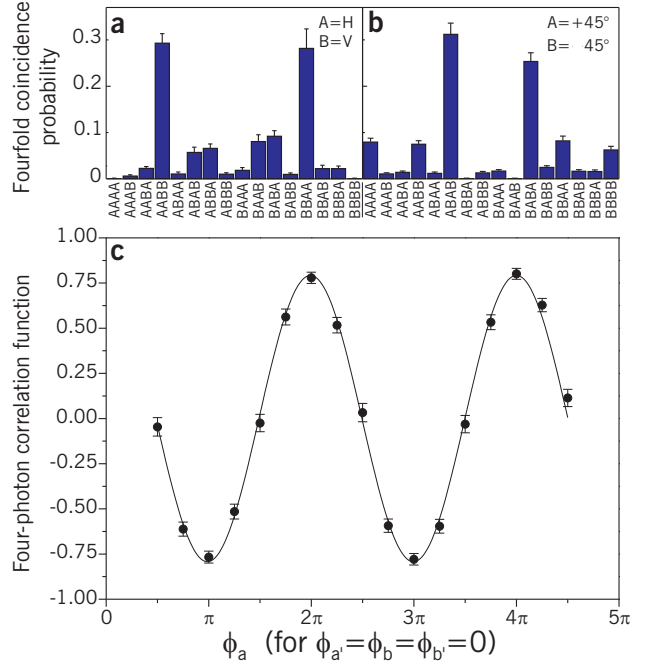,width=8.5cm}
\end{center}
\caption{\label{coincifig}
Fourfold coincidence probabilities corresponding to a
detection of one photon in each of  the four polarization
analyzers, oriented in (a) the H/V basis, and (b) the
$\pm45^\circ$ basis. (c) Four-photon polarization correlation with
the detection basis of an observer in mode $a$ varying from
$45^\circ$ linear at $\phi_a=0$ to left circular, $-45^\circ$
linear and right circular polarization, while observers in $a',b,
b'$ analyze in the $\pm45^\circ$ basis. The solid line shows a
sinusoidal fit to the experimental results  with a visibility of
$79.3\pm1.4\%$.}
\end{figure}

Fig.~\ref{coincifig}a shows the 16 possible fourfold coincidence probabilities
for detecting one photon in each of the four outputs of the beam
splitters, with all four polarization analyzers oriented along
$H/V$. The rates of the $HHVV$ and the $VVHH$ events are in very
good agreement with the state in equation (\ref{state}) and, within errors,
equal to the sum of all events where the two photons detected in
arms $a$ and $a'$, or in arms $b$ and $b'$, have orthogonal
polarization. The four-photon state $\ket{\Psi^{(4)}}$ exhibits the mentioned
invariance under identical change of the four detection bases.
 Fig.~\ref{coincifig}b shows the four-photon
coincidence probability when analyzed along $+45^\circ/-45^\circ$
linear polarization. Again, one observes two types of
coincidences, the GHZ part, and the fourfold coincidences due to the EPR
pairs with average rates lower by a factor of four.
 Integration times were 5~h, and 17.5~h,
respectively, with four-fold coincidence rates between 300 and 100
per hour, varying mainly due to drifts of the fiber coupling.

As a first step in the characterization of entangled states it is
customary to analyze the correlations between measurement results.
For this purpose, polarization measurements corresponding to
dichotomic observables with eigenvectors $\ket{l_{x}, \phi_x} =
\sqrt{1/2}(\ket{V}_{x}+l_{x} e^{-i\phi_x}\ket{H}_{x})$ and
eigenvalues $l_{x}=\pm1$ are performed by the observation
stations in the four modes ($x=a,a',b,b'$). The theoretical
prediction for the correlation function defined as the expectation
value of the product of the four local results is given by
\cite{proposal}
\begin{eqnarray}
E_{QM}(\phi_{a},\phi_{a'},\phi_{b},\phi_{b'})&=&{2\over3}
\cos(\phi_{a}+\phi_{a'}-\phi_{b}-\phi_{b'}) \label{corrfunc}  \\
&&+{1\over3}\cos{(\phi_{a}-\phi_{a'})} \cos{(\phi_{b}-\phi_{b'})}.
\nonumber
\end{eqnarray}

The experimental value of the correlation function can be obtained
from the 16 four-photon coincidence rates,  as shown in Fig.~\ref{coincifig}a
and b, via 
\begin{eqnarray}
&&E(\phi_{a},\phi_{a'},\phi_{b},\phi_{b'})  \nonumber \\
&&=\sum_{l_a,l_{a'},\atop l_b,l_{b'}=\pm1}l_al_{a'}l_bl_{b'}p_{l_a,l_{a'},l_b,l_{b'}}(\phi_{a},\phi_{a'},\phi_{b},\phi_{b'}).
\end{eqnarray}
Therein, the four-photon probabilities $p_{l_a,l_{a'},l_b,l_{b'}}$ are given
by 
$$p_{l_a,l_{a'},l_b,l_{b'}}(\phi_{a},\phi_{a'},\phi_{b},\phi_{b'})
= c_{l_a,l_{a'},l_b,l_{b'}}/\sum c\;,$$
where $c_{l_a,l_{a'},l_b,l_{b'}}$ is the number of recorded four-fold events
at the detectors specified by the indices (for the specific settings), and 
the sum is the total number of relevant four-fold
events. Fig.~\ref{coincifig}c shows the dependence of the correlation function
on the angle $\phi_a$, for the other analyzers fixed at angles
$\phi_{a'}=\phi_{b}=\phi_{b'}=0$, corresponding to $45^\circ$
linear polarization. 
For its visibility, which here is equal to the maximal absolute value of the
correlation function (eq.~(3)), we obtain a value of $V=79.3\%\pm1.4\%$. 
It serves as a measure for the quality of our
state preparation, and largely depends on the ratio between the
spectral bandwidth of the down-conversion pump light and the
spectral width of the detected photons \cite{ZZW}.

Note that the analysis angles giving perfect correlations of
$\ket{\Psi^{(4)}}$ are different from those for a four-photon
GHZ state. Due to the EPR contributions, this state cannot be used
in a GHZ-type argument refuting local hidden variable models of
quantum mechanics. However, the invariance mentioned earlier enables
perfect correlations for all possible sets of common analysis
directions, a feature which does not hold for GHZ states, but which is of
great importance for several quantum communication schemes.

The contribution of the product of the EPR states also leads to a
different nature of the four-photon entanglement. The seemingly
innocent question of how much entanglement is in the state
$\ket{\Psi^{(4)}}$ cannot be answered for the moment,
 because clear measures of multi-particle entanglement are still missing.
 Contrary
to the case for two particles, multi-particle entanglement can be
classified from numerous viewpoints under discussion
\cite{quantent}. Keeping in mind possible applications for
multi-party quantum cryptography and secret sharing, we analyze
the entanglement of the state in terms of violation of a
(non-conventional) Bell inequality.

One can write down a {\em single} Bell inequality which summarizes
all possible local realistic constraints on the correlation
function for the case of each local observer measuring the
polarizations along {\em two} alternative directions
\cite{proposal,WERNER,ZUKOWSKI}. Let us introduce a shorthand
notation, $E(\phi_{a}^k, \phi_{a'}^l, \phi_{b}^m, \phi_{b'}^n)$,
for the correlation functions deduced from the observed count
rates for the full set of 16 local directions, with
$k,l,m,n=1,2$ denoting which of the two alternative phase settings
was chosen at the local observation station measuring in arm
$x=a,a',b,b'$. The generalized Bell inequality  reads \cite{ZUKOWSKI}

\begin{eqnarray}
&&S^{(4)}=   \label{Bell}\\ && ={1\over2^4}\sum\limits_{s_x=\pm1 \atop
{x=a,a',b,b'}}\left|
  \sum\limits_{k,l,m,n \atop =1,2}s_a^ks_{a'}^l s_b^m s_{b'}^n\, E(\phi_a^k,
\phi_{a'}^l,
  \phi_b^m, \phi_{b'}^n) \right| \le 1. \nonumber
\end{eqnarray}

The maximal violation of this inequality for $\ket{\Psi^{(4)}}$ is
obtained when three observers (one in each mode $x=a',b,b'$) perform
polarization 
analysis along $\phi^{1,2}_x=\pm \pi/4$, and the observer in
mode $a$ chooses between $\phi_{a}^1=0$ or $\phi_{a}^2=\pi/2$.
Then the quantum prediction is as high as $S^{(4)}_{QM}=1.886$, and
results in a violation of the above inequality whenever the
correlation function implied by the studied state
$\ket{\Psi^{(4)}}$ has visibility greater 53\%\cite{proposal}. In
comparison, for a four-photon GHZ state one obtains
$S^{(4)}_{QM}=\sqrt{8}$ and a critical visibility of
$1/\sqrt{8}\approx35\%$. The visibility requirement for an
experimental violation of the inequality (\ref{Bell})  is
therefore more demanding for $\ket{\Psi^{(4)}}$. But since it is
much simpler to generate $\ket{\Psi^{(4)}}$, an experimental
violation of the local realistic condition (\ref{Bell}) becomes
feasible.

\begin{figure}
\begin{center}
\epsfig{file=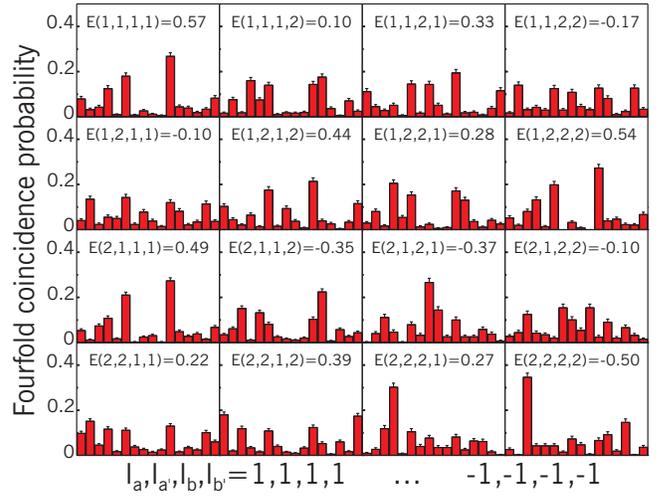,width=8.5cm}
\end{center}
\caption{\label{bellfig}
Fourfold coincidence probabilities (raw data) for a four-particle
test of local realistic theories. For the sixteen settings of the
analyzer phases $\phi_{a}, \phi_{a'}, \phi_{b}, \phi_{b'}$, count rates {\em
  not} corrected for detection efficiencies are used to evaluate a generalized
Bell inequality (\ref{Bell}), leading to $S^{(4)}=1.301\pm0.040$. This clearly
exceeds the bound of 1 demanded by local realistic theories.
Acquisition time for the individual frames was ranging from 2.75~h
to 4.25~h, with around 150 fourfold coincidence events per hour.}
\end{figure}

Fig.~\ref{bellfig} shows all 256 fourfold coincidence probabilities necessary
for such an analysis. They were recorded in blocks of 16 coincidence
rates, each corresponding to one of the 16 phase settings
appearing in the inequality (\ref{Bell}). Integration times were ranging from 2.75 to 4.75 hours per frame. The whole measurement including
regular realignment etc. took about four days. For evaluating the
generalized Bell-inequality we used the {\it raw} data without any
correction for background and relative collection or detection efficiency,
etc. The resulting value $S^{(4)}=1.301 \pm 0.040$ strongly violates the
boundary for local realistic theories and confirms the
entanglement of $\ket{\Psi^{(4)}}$.

Perfect correlations and the violation of a Bell-inequality are
the key ingredients of entanglement based quantum cryptography
\cite{Ekert91}, which with the aid of $\ket{\Psi^{(4)}}$ can be now expanded
for multi-party quantum communication. Similar to a four-observer Bell
experiment, parties A', B, and B' (observing photons in mode $a'$,  
$b$ and $b'$)  switch between analysis angles $\phi_x=\pm\pi/4$, while 
party A observing photons in mode $a$ switches $\phi_a$ between $0$ or
$\pi/2$, and with a certain 
probability to $\pi/4$. After a number of quartets are
registered, party A announces the times when having analyzed along $0$ or
$\pi/2$. For these detection quartets the other parties publicly
announce their settings and results, which now can be used to
evaluate $S^{(4)}$. The degree of violation of the Bell inequality
(\ref{Bell}) is a measure of the security of the key exchange. Any
eavesdropper attack on any of the four quantum channels would
reduce the entanglement and thus the violation of inequality
(\ref{Bell}). Thus, the four parties can assume that the
remaining instances have been securely transmitted. Since in these cases
party A measured along $\phi_a=\pi/4$, perfect correlations exist
between the four measurement results according to equation (\ref{corrfunc}) and
enable the four parties 
to obtain a random, secure key \cite{BBM}.

Two ways to obtain a secure key can be formulated. Firstly, and
similarly to recently proposed schemes of secret sharing
\cite{secrshare}, the parties could cooperate such that two of the
four reveal their settings and results to the other two. Relying
on the perfect correlations of the state, each of the two
remaining parties can infer the result of the other and thus can
evaluate a secure key only known to these two parties. Secondly,
the measurement results can be used to {\em distribute} a key
to three of the four parties. This can be achieved if, e.g., 
parties A and A' cooperate (forming now a single party A*) and
compare their measurement results. The instances where they
obtain the same results, are only due to the GHZ contribution of
$\ket{\Psi^{(4)}}$. For these cases the correlations of the
GHZ state allow now the three parties A*, B, and B' to create a
common secret key  \cite{GHZremark}.

In summary, bosonic-type interference can be utilized to produce
multi-photon entanglement directly from spontaneous parametric
down-conversion.  Without interferometric setups we could
demonstrate the correlations between measurement results of four
observers and the violation of a generalized four-photon Bell
inequality. The high visibility of the quantum correlations and the ease of
operation of our 
source show its potential for multi-party quantum communication
applications like quantum secret sharing, three-party key
distribution or for quantum telecloning.

MZ was supported by Komitet Bada{\'n} Naukowych, grant No. 5 P03B 088 20. This
work was supported by the EU-Project QuComm (IST-FET-10033), and the Deutsche
Forschungsgemeinschaft (We 2451/1-2).



\begin{references}
\bibitem{otherdegfreed}This also holds for other degrees of freedom
causing mode, frequency or time--energy entanglement.
The generic effect also leads to entanglement
between a higher number of photons. Interferometrically enhanced
down-conversion \cite{Lamas01} may allow access 
to states with more than four photons in the future.
\bibitem{GHZremark}Of course, a similar argument holds when measuring a
  GHZ state right away. However, since the GHZ state lacks the particular
  symmetry of $\ket{\Psi^{(4)}}$ one has to switch between different
  bases for key generation or for security check, thus reducing the
  key rate significantly.
\end{references}
\end{document}